\documentstyle[12pt,a41]{article}

\begin{document}
%\setlength{\baselineskip}{0.515cm}
%\sloppy

\thispagestyle{empty}

%\mbox{}
\begin{flushleft}
DESY  98-137 \hfill 
{\tt hep-ph/9809411}\\
September 1998
\end{flushleft}

\setcounter{page}{0}

\mbox{}
\vspace*{\fill}
\begin{center}
{\LARGE\bf Second order QCD corrections to}

\vspace{2mm}
{\LARGE\bf the forward-backward asymmetry in ${\bf e^+~e^-}$-collisions}

\vspace*{20mm}
\large
{V. Ravindran and W.L. van Neerven \footnote{On leave of absence from 
Instituut-Lorentz, University of Leiden,P.O. Box 9506, 2300 RA Leiden,\\ 
The Netherlands
.}}
\\

\vspace{2em}

\normalsize
%{\it DESY--Zeuthen, Platanenallee 6,}\\
%{\it D--15735 Zeuthen,  Germany}
{\it DESY-Zeuthen, Platanenallee 6, D-15738 Zeuthen, Germany}%

\vspace*{\fill}
%\vspace{4cm}
%\today
\end{center}
\begin{abstract}
\noindent
We will present the result of an analytical calculation of the second
order contribution to the forward-backward asymmetry $A_{\rm FB}^{\rm H}$ and 
the shape constant $a^{\rm H}$ for heavy flavour production in 
$e^+~e^-$-collisions. The calculation has been carried out by assuming
that the quark mass is equal to zero. This is a reasonable approximation for 
the exact second order correction for charm and bottom quark production at LEP
energies but not for top production at future linear colliders. Our
result for $A_{\rm FB}^{\rm H}$ is a factor 2.6 (charm) and 4.7 (bottom)
larger than obtained by a numerical calculation performed earlier in 
the literature. We study the effect of the second order corrections
on the above parameters including their dependence on the renormalization scale.
Further we make a comparison between the fixed pole mass and the running mass
approach.
\end{abstract}
%\vspace*{1cm}

\vspace*{\fill}
Experiments carried out at electron positron colliders like LEP and LSD 
have provided us with a wealth of information about the constants \cite{abc}
appearing in the standard model of the electroweak and strong interactions.
One among them is given by the electroweak mixing angle defined by $\theta_W$
which can be very accurately extracted from the forward-backward asymmetry
in heavy flavour production in particular when the flavour
is represented by the bottom quark \cite{dkz}. Recently this quantity
is obtained for the charm quark \cite{bara} and in the future one also
hopes to measure it for the top quark at the
large linear $e^+~e^-$-collider (see e.g. \cite{dhk}). 
The forward-backward asymmetry
is extracted from the differential cross section given by
\begin{eqnarray}
\label{eq1}
  \frac{d\sigma^{\rm H}(Q^2)}{d\cos\theta} &=& \frac{3}{8}
(1+\cos^2\theta)\left [\sigma_{\rm VV}(Q^2) f_T^v(\rho) 
+ \sigma_{\rm AA}(Q^2) f_T^ a(\rho) \right ]
\nonumber\\[2ex]
&&  + \frac{3}{4}\sin^2\theta \left [ \sigma_{\rm VV}(Q^2) f_L^v(\rho) +
\sigma_{\rm AA} (Q^2) f_L^a(\rho) \right ]
 + \frac{3}{4}\cos\theta \left [\sigma_{\rm VA}(Q^2) f_A^a(\rho) \right ] \,,
\end{eqnarray}
where $\theta$ is the angle between the outgoing quark H and the
incoming electron. The CM energy is denoted by Q and the scaling variable $\rho$
is defined by $\rho=4m^2/Q^2$ where $m$ stands for the quark mass.
Notice that the form of the cross section above is correct if only final
state corrections are present which is the case for QCD investigated in this
paper. For electroweak corrections (see \cite{bard}, \cite{bhm})
which occur in the initial as well as in the final state, including interference 
terms, the above formula has to be modified.
The Born cross sections for quark final states appearing in Eq. (\ref{eq1}) 
can be written as
\begin{eqnarray}
\label{eq2}
  \sigma_{VV}(Q^2) &=& \frac{4\pi\alpha^2}{3Q^2}N\,\left [e_\ell
^2 e_q^2 +
 \frac{2Q^2(Q^2-M_Z^2)}{\left|Z(Q^2)\right|^2}\,e_\ell e_q C_{V,\ell} C_{V,q} 
\right.
\nonumber\\
&& \left. + \frac{(Q^2)^2}{\left|Z(Q^2)\right|^2} 
   \left (C_{V,\ell}^2 + C_{A,\ell}^2\right )C_{V,q}^2 \right ] \,,\\
\label{eq3}
  \sigma_{AA}(Q^2) &=& \frac{4\pi\alpha^2}{3Q^2}N\,\left [
  \frac{(Q^2)^2}{\left|Z(Q^2)\right|^2}
  \left (C_{V,\ell}^2 + C_{A,\ell}^2\right ) C_{A,q}^2 \right ] \,,\\
\label{eq4}
\sigma_{VA}(Q^2) &=& \frac{4\pi\alpha^2}{3Q^2}N\,\left [\frac{2Q^2(Q^2-M_Z^2)}
  {\left|Z(Q^2)\right|^2}\,e_\ell e_q C_{A,\ell} C_{A,q}
+  4\frac{(Q^2)^2}{\left|Z(Q^2)\right|^2}\,C_{A,\ell} C_{A,q} C_{V,\ell} 
C_{V,q} \right ] \,,
\end{eqnarray}
where $N$ denotes the number of colours in the case of the gauge group
$SU(N)$ (in QCD one has $N=3$). Furthermore
in the expressions above we adopt for the Z-propagator the energy independent
width approximation
\begin{eqnarray}
\label{eq5}
  Z(Q^2) = Q^2 - M_Z^2 + iM_Z\Gamma_Z \,.
\end{eqnarray}
The charges of the lepton and the quark are given by $e_\ell$ and $e_q$
respectively and 
the angle $\theta_W$ defined at the beginning appears in the electroweak
constants as follows.
\begin{eqnarray}
\label{eq6}
  \begin{array}{ll}
    C_{A,\ell} = \displaystyle\frac{1}{2\sin2\theta_W}, &
    C_{V,\ell} = -C_{A,\ell}\,(1-4\sin^2\theta_W),\\[2ex]
    C_{A,u} = -C_{A,d} = -C_{A,\ell}, & \\[2ex]
    C_{V,u} = C_{A,\ell}\,\displaystyle(1-\frac{8}{3}\sin^2\theta_W), &
    C_{V,d} = -C_{A,\ell}\,\displaystyle(1-\frac{4}{3}\sin^2\theta_W) \,.
  \end{array}
\end{eqnarray}
The functions $f_k^l$ ($k=T,L,A;~l=v,a$) in Eq. (\ref{eq1}) can be
computed order by order in perturbative QCD from the non-singlet quark 
coefficient function ${\cal C}_{k,q}^{l,\rm NS}$ as follows
\begin{eqnarray}
\label{eq7}
f_k^l(\rho)=\int_{\sqrt \rho}^1 dx \,{\cal C}_{k,q}^{l,\rm NS}
(x,\rho,\frac{Q^2}{\mu^2}) \quad \mbox{with} \quad \rho = \frac{4m^2}{Q^2}\,,
\end{eqnarray}
where $\mu$ stands for the factorization as well as the renormalization scale.
The quark coefficient function  appears in the fragmentation function 
$F_k(x,Q^2)$ with $x=2p.q/Q^2$ where $p$ is the momentum of the outgoing hadron
which originates from the quark. These fragmentation functions
describe the production of the quark and its subsequent decay into a hadron.
The forward backward asymmetry, denoted by $A_{\rm FB}^{\rm H}$, appears when
we divide the expression in Eq. (\ref{eq1}) by the total cross section. The  
ratio can then be expressed in the following way
\begin{eqnarray}
\label{eq8}
\frac{1}{\sigma_{\rm tot}^{\rm H}(Q^2)} \frac{d\sigma^{\rm H}(Q^2)}{d\cos\theta}
= \frac{3}{8} \left (\frac{4}{3 + a^{\rm H}(Q^2)} \right )
\left (1 + a^{\rm H}(Q^2) \cos^2\theta \right )
 + A_{\rm FB}^{\rm H}(Q^2) \cos \theta \,,
\end{eqnarray}
with
\begin{eqnarray}
\label{eq9}
A_{\rm FB}^{\rm H}(Q^2)=\frac{3}{4} \frac{\sigma_{\rm VA}(Q^2)\,f_A^a(\rho) }
{\sigma_{\rm tot}^{\rm H}(Q^2)} \,.
\end{eqnarray}
Further the shape coefficient $a^{\rm H}$ is defined by
\begin{eqnarray}
\label{eq10}
a^{\rm H}(Q^2)=\frac{\sigma_{\rm VV}(Q^2) \Big [f_T^v(\rho)
-2f_L^v(\rho) \Big ]+
\sigma_{\rm AA}(Q^2) \Big [f_T^a(\rho)-2f_L^a(\rho)\Big ]}
{\sigma_{\rm VV}(Q^2)\Big [f_T^v(\rho)+2f_L^v(\rho)\Big ]+
\sigma_{\rm AA}(Q^2)\Big [f_T^a(\rho)+2f_L^a(\rho)\Big ]} \,,
\end{eqnarray}
and the total cross section for heavy flavour production is equal to
\begin{eqnarray}
\label{eq11}
\sigma_{\rm tot}^{\rm H}(Q^2)=\sigma_{\rm VV}(Q^2)\Big [f_T^v(\rho) 
+ f_L^v(\rho)\Big ]
+ \sigma_{\rm AA}(Q^2)\Big [f_T^a(\rho) + f_L^a(\rho)\Big ] \,.
\end{eqnarray}
The functions $f_k^l$ can be expanded in the strong coupling constant
$\alpha_s$ as follows
\begin{eqnarray}
\label{eq12}
f_k^l(\rho) = \sum_{n=0}^{\infty} \left (\frac{\alpha_s(\mu)}{4\pi} \right )^n
f_k^{l,(n)}(\rho) \,.
\end{eqnarray}
The lowest order contributions corresponding to the Born reaction 
\begin{eqnarray}
\label{eq13}
V \rightarrow {\rm H} + {\overline {\rm H}} \,,
\end{eqnarray}
with $V= \gamma, Z$ are given by
\begin{eqnarray}
\label{eq14}
f_T^{v,(0)}(\rho) &=& \sqrt{1- \rho} \qquad f_L^{v,(0)}(\rho)
=\frac{\rho}{2}
\sqrt{1- \rho} \,,
\nonumber\\[2ex]
f_T^{a,(0)}(\rho) &=& (1- \rho)^{3/2} \qquad f_L^{a,(0)}(\rho)=0 \,,
\nonumber\\[2ex]
f_A^{a,(0)}(\rho) &=& 1- \rho \,,
\end{eqnarray}
where $\rho$ is defined below Eq. (\ref{eq1}). The next-to-leading order (NLO)
contributions originate from the one-loop virtual corrections to the Born 
reaction
(\ref{eq13}) and the gluon bremsstrahlungs process
\begin{eqnarray}
\label{eq15}
V \rightarrow {\rm H} + {\overline {\rm H}} + g \,.
\end{eqnarray}
The NLO contributions have been calculated by several groups in the literature  
for the case $m\not =0$ (see \cite{laer}-\cite{stol}). 
However there were some discrepancies
between the results so that we decided to calculate them again using a different
method. Here they are derived from the order $\alpha_s$ contribution to the 
coefficient functions for heavy quarks which appears in the integrand of 
Eq. (\ref{eq7}). Our computations lead to the
same answers as given in appendix A of \cite{nawe}. After
performing the integral in  Eq. (\ref{eq7}) we obtain the following results
for $k=T,L$
\begin{eqnarray}
\label{eq16}
f_T^{v,(1)}(\rho)&=&C_F\Big [\frac{1}{2}\rho(1 + \rho)F_1(t)
+\sqrt \rho(1 - 3\rho)
F_2(t)+(32 - \frac{39}{2}\rho - \frac{7}{2} \rho^2) {\rm Li}_2(t)
\nonumber\\[2ex]
&& +(16 - 10 \rho - 2\rho^2)F_3(t)+2 \sqrt {1-\rho}\,F_4(t)
+(8 - 6\rho - 2\rho^2)\ln(t) \ln(1+t)
\nonumber\\[2ex]
&& +(-12 + 9 \rho - \frac{5}{4} \rho^2)\ln(t)
+ \sqrt {1-\rho}\, (1 + \frac{13}{2} \rho) \Big ] \,,\\
\label{eq17}
f_L^{v,(1)}(\rho)&=&C_F \Big [-\frac{1}{2}\rho(1 + \rho)F_1(t)
-\sqrt \rho(1 - 3\rho)
F_2(t)+(\frac{39}{2}\rho - \frac{9}{2} \rho^2) {\rm Li}_2(t)
\nonumber\\[2ex]
&& +( 10 \rho - 2\rho^2)F_3(t)+ \rho \sqrt {1-\rho}\,F_4(t)+6\rho\ln(t) \ln(1+t)
\nonumber\\[2ex]
&& +(-7 \rho + 3 \rho^2)\ln(t) + 2 (1 - \rho)^{3/2} \Big ] \,,\\
\label{eq18}
f_T^{a,(1)}(\rho)&=&C_F \Big [\frac{1}{2}\rho(1 + 2 \rho)F_1(t)
+\sqrt \rho(1 - 4\rho)
F_2(t)+(32 - \frac{103}{2}\rho + 9 \rho^2) {\rm Li}_2(t)
\nonumber\\[2ex]
&& +(16 - 26 \rho + 4\rho^2)F_3(t)+2 (1 - \rho)^{3/2}F_4(t)
+(8 - 14\rho )\ln(t) \ln(1+t)
\nonumber\\[2ex]
&& +(-12 + 15 \rho - \frac{9}{4} \rho^2)\ln(t)
+ \sqrt {1-\rho}\, (1 + \frac{1}{2} \rho) \Big ] \,,\\
\label{eq19}
f_L^{a,(1)}(\rho)&=&C_F \Big [-\frac{1}{2}\rho(1 + 2 \rho)
\Big (F_1(t)- 4 F_3(t)
 - 7{\rm Li}_2(t) - 4 \ln(t) \ln(1+t)\Big )
\nonumber\\[2ex]
&&-\sqrt \rho(1 - 4\rho) F_2(t) + (-4 \rho + \rho^2 - \frac{3}{8} \rho^3)\ln(t)
+ \sqrt {1-\rho}\, (2 - \frac{19}{2} \rho + \frac{3}{4} \rho^2) \Big ] \,,
\end{eqnarray}
with
\begin{eqnarray}
\label{eq20}
t=\frac{1-\sqrt{1-\rho}}{1+\sqrt{1-\rho}} \,.
\end{eqnarray}
Further the colour factor $C_F$ is given by $C_F=(N^2-1)/2N$.
The functions $F_i(t)$ appearing above are defined by
\begin{eqnarray}
\label{eq21}
F_1(t)&=&{\rm Li}_2(t^3)+4\zeta(2)+\frac{1}{2}\ln^2(t)+3\ln(t)\ln(1+t+t^2)\,,\\
\label{eq22}
F_2(t)&=&{\rm Li}_2(-t^{3/2})-{\rm Li}_2(t^{3/2})+{\rm Li}_2(-t^{1/2})
-{\rm Li}_2(t^{1/2}) + 3 \zeta(2) + 2\ln(t)\ln(1 + \sqrt t) 
\nonumber\\
&& -2\ln(t)\ln(1 - \sqrt t) + \frac{3}{2}\ln(t)\ln(1+t-\sqrt t)
 -\frac{3}{2}\ln(t)\ln(1+t+\sqrt t) \,,\\
\label{eq23}
 F_3(t)&=&{\rm Li}_2(-t) + \ln(t)\ln(1 - t) \,,\\
\label{eq24}
F_4(t)&=&6 \ln(t) - 8\ln(1 - t) - 4\ln(1 + t) \,,
\end{eqnarray}
where $\zeta(n)$, which appears for $n=2,3$ in the formulae of this paper,
represents the Riemann $\zeta$-function and ${\rm Li}_2(x)$ denotes
the dilogarithm.
Using Eqs. (\ref{eq16})-(\ref{eq19}) one can check that the order $\alpha_s$
contribution to the total cross section in (\ref{eq11}) is in agreement with
the literature \cite{stn} (for the vector part see also \cite{kasa}).
Integration of the asymmetry quark coefficient function provides us with the
result
\begin{eqnarray}
\label{eq25}
 f_A^{a,(1)}(\rho)&=&C_F \Big [-4 (2 -  \rho)\sqrt{1 - \rho}\,G_1(t)
+2(4 - 5 \rho)G_2(t) + 8 \ln(1 + t - \sqrt t) 
\nonumber\\[2ex]
&&  -8(1-\rho)\left (\ln(1 + t) + 2 \ln(1 - \sqrt t)\right ) 
+ \left ( 4 (1 - 2 \rho ) + 2 \sqrt{1 - \rho}\,( - 2 + 3 \rho)\right )\ln(t)
\nonumber\\[2ex]
&&  + 4(\rho - \sqrt \rho) \Big ] \,,
\end{eqnarray}
which involves the following functions
\begin{eqnarray}
\label{eq26}
 G_1(t)&=&{\rm Li}_2(-t^{3/2}) - 3 {\rm Li}_2(-t^{1/2}) - 4 {\rm Li}_2(t^{1/2})
- {\rm Li}_2(-t) - \frac{1}{2} \zeta(2) - \frac{1}{8} \ln^2(t) \,,\\
\label{eq27}
 G_2(t)&=&{\rm Li}_2\left (\frac{\sqrt t }{1 + t}\right) -\frac{1}{2}\zeta(2)
- \frac{1}{2} \ln(t)\ln(1 + t) +  \frac{1}{2} \ln^2(1 + t) 
- \frac{1}{8} \ln^2(t) \,.
\end{eqnarray}
The functions $f_k^{l,(1)}$ ($k=T,L;~l=v,a$) are related to the functions 
$H_2$, and $H_6$ presented in Eq. (15) and appendix A of \cite{abl} and we
agree with their result. The same holds for $f_A^{a,(1)}$ which is
proportional to $H_5$ in the reference above. There is also agreement with
the calculation in \cite{stol}. The comparison is made by expanding the 
functions above and those in \cite{abl} up to seven powers in $\rho$. The 
next-to-next-to-leading
order (NNLO) contributions come from the following processes. First one has to
compute the two-loop vertex corrections to the Born process (\ref{eq13}) and
the one-loop corrections to (\ref{eq15}). Second one has to add the radiative
corrections due to the following reactions
\begin{eqnarray}
\label{eq28}
V \rightarrow {\rm H} + {\overline {\rm H}} + g + g \,,
\end{eqnarray}
\begin{eqnarray}
\label{eq29}
V \rightarrow {\rm H} + {\overline {\rm H}}+{\rm H} + {\overline {\rm H}}\,,
\end{eqnarray}
\begin{eqnarray}
\label{eq30}
V \rightarrow {\rm H} + {\overline {\rm H}} + q + {\overline q} \,,
\end{eqnarray}
where q denotes the light quarks. The results for $f_k^{l,(2)}$ presented 
below are computed for the contributions where
the vector boson V is always coupled to the heavy quark H so that the light 
quarks in Eq. (\ref{eq30}) are only produced via fermion pair production 
emerging from gluon splitting.
Besides these contributions there are other ones which have been treated in 
\cite{alla}. The latter consist of all one- and two-loop vertex corrections
which contain the triangular quark-loop graphs. Following the notation in
\cite{alla} their contribution
to the forward-backward symmetry will be denoted by $F_{QCD}^{3-jet}$ 
and $F_{QCD}^{2-jet}$ respectively. They only show up if the quarks
are massive and are coupled to the Z-boson via the axial-vector vertex. Notice
that one has to sum over all members of one quark family in order to cancel
the anomaly. Adopting the mass assignment in \cite{alla} we take the top to 
be massive and put the other quark masses, including that of the bottom, equal 
to zero.
Further in \cite{alla} one has included all terms originating from reaction
(\ref{eq30}) where diagrams with light quarks attached to the vector boson V
interfere with those describing the coupling of the heavy quarks to the
vector boson. Notice that this contribution denoted in 
\cite{alla} by $F_{QCD}^{\rm F}$ vanishes if
all quarks including H are taken to be massless provided one sums over all 
members in one family. However we will omit that part of reaction
(\ref{eq30}) where the heavy quarks are produced via gluon splitting and the
light quarks q are coupled to the vector boson V. This contribution denoted
by $F_{QCD}^{\rm Branco}$ in \cite{alla} needs a cut on the invariant
mass of the heavy flavour pair and it was computed for the first time in 
\cite{bns}). 
Finally notice that these additional contributions, denoted by $F_{QCD}$ above,
only show up in order $\alpha_s^2$. Moreover if we put $m=0$ they only appear
in the forward-backward asymmetry (\ref{eq9}) but cancel between 
numerator and denominator in the shape coefficient (\ref{eq10}).
The results for $f_k^{l,(2)}$ follow from the 
transverse and longitudinal coefficient functions calculated in \cite{rijk1}
and the asymmetry coefficient function computed in \cite{rijk2}. Because of the 
complexity of the calculation of these functions the heavy quark mass was taken 
to be zero.
This approximation is good for the charm and bottom quark but not for the top
quark as we will see below. Substituting these coefficient functions in the
integrand of Eq. (\ref{eq7}) we obtain
\begin{eqnarray}
\label{eq31}
 f_T^{v,(2)}=f_T^{a,(2)}&=&C_F^2 \left \{ \frac{7}{2} \right \} 
+ C_AC_F \left \{-\frac{11}{3}
\ln \left ( \frac{Q^2}{\mu^2}\right )+\frac{347}{18}-44\zeta(3)\right \}
\nonumber\\
&&  +n_f C_F T_f \left \{\frac{4}{3}\ln \left ( \frac{Q^2}{\mu^2}\right ) 
- \frac{62}{9} + 16 \zeta(3) \right \} \,,\\
\label{eq32}
f_L^{v,(2)}=f_L^{a,(2)}&=&C_F^2\left \{-5 \right \}+C_AC_F\left \{-\frac{22}{3}
\ln \left ( \frac{Q^2}{\mu^2}\right ) + \frac{380}{9} \right \}
\nonumber\\
&& +n_f C_F T_f \left \{\frac{8}{3}\ln \left ( \frac{Q^2}{\mu^2}\right ) 
- \frac{136}{9}\right \} \,,\\
\label{eq33}
f_A^{a,(2)}&=&C_A C_F \left \{- 44 \zeta(3)\right  \} 
+ n_f C_F T_f \left \{ 16 \zeta(3) \right \} \,.
\end{eqnarray}
Here the colour factors are given by $C_A=N$ and $T_f=1/2$ (for $C_F$
see below Eq. (\ref{eq20})). Further $n_f$ 
denotes the number of light flavours which originate from process (\ref{eq30}).
Finally $\mu$ appearing in the strong coupling constant $\alpha_s$ and
the logarithms in Eqs.(\ref{eq31}), (\ref{eq32}) represents the
renormalization scale. Notice that the coefficient of the logarithm is
proportional to the lowest order coefficient of the $\beta$-function. The
same holds for $\zeta(3)$ in Eqs. (\ref{eq31}) and (\ref{eq33}). 
The logarithm does not appear in $f_A^{a,(2)}$
because $f_A^{a,(1)}=0$ at $m=0$. Since the mass is equal to zero
there is no distinction anymore between $f_k^{v,(2)}$ and $f_k^{a,(2)}$
($k=T,L$) unlike in the case for the first order corrections
in Eqs. (\ref{eq16})-(\ref{eq19}) where the heavy quark was taken to be massive.
Furthermore one can check that substitution of Eqs. (\ref{eq31}), (\ref{eq32})
into (\ref{eq11}) provides us with the order $\alpha_s^2$ contribution to the
total cross section which is in agreement with the results obtained in 
\cite{ckt}.
For zero mass quarks the forward-backward asymmetry becomes equal to
\begin{eqnarray}
\label{eq34}
A_{\rm FB}^{\rm H}(Q^2)&=& A_{\rm FB}^{{\rm H},(0)}(Q^2)\left [ 1 - 
\frac{\alpha_s(\mu)}{4\pi} C_F\left \{ 3 \right \} + 
\left (\frac{\alpha_s(\mu)}{4\pi}\right )^2\left\{
C_F^2\left(\frac{21}{2}\right) \right. \right.
\nonumber\\[2ex]
&& \left. \left. 
+ C_AC_F \left ( 11 \ln \left(\frac{Q^2}{\mu^2}\right) - \frac{123}{2} \right ) 
+ n_fC_F T_F \left (-4 \ln \left(\frac{Q^2}{\mu^2}\right) 
+ 22 \right ) \right \} \right ] \,.
\end{eqnarray}
For the discussion below and the notation often used in the literature
(see \cite{dkz})
it is convenient to write $A_{\rm FB}^{\rm H}$ in the following way
\begin{eqnarray}
\label{eq35}
A_{\rm FB}^{\rm H}(Q^2)
= A_{\rm FB}^{{\rm H},(0)}(Q^2) \left [ 1 - \frac{\alpha_s(\mu)}{\pi}c_1
- \left (\frac{\alpha_s(\mu)}{\pi} \right )^2 c_2 \right ] \,.
\end{eqnarray}
The order $\alpha_s^2$ contribution presented above has been also calculated in 
\cite{alla} but then in a numerical way. Unfortunately we get a different 
answer.
First we disagree with the statement above Eq. (48) in \cite{alla} that reaction
(\ref{eq29}) does not contribute to $A_{\rm FB}^{\rm H}$. For the latter
we obtain the following contribution to $c_2$, denoted by $\Delta^{(1)} c_2$,
which equals
\begin{eqnarray}
\label{eq36}
\Delta^{(1)} c_2= - (C_F^2-\frac{1}{2} C_AC_F)\frac{1}{16} 
\left [ -\frac{19}{2} +6 \zeta(2) + 8 \zeta(3) \right ] = 0.14 \,.
\end{eqnarray}
The result for reactions (\ref{eq28}) and (\ref{eq30}), including the virtual 
corrections, can be obtained by subtracting
Eq. (\ref{eq36}) from Eq. (\ref{eq34}). Using the notation in Eq. (45) 
\cite{alla} and choosing $n_f=5$ we find the following contributions
\begin{eqnarray}
\label{eq37}
 c_1 = \frac{3}{4} C_F && \qquad  \Delta^{(2)} c_2= - \frac{1}{4}C_F
\left [ \frac{9}{4}C_F + CC_F + N N_C + T T_R \right ] \,,
\nonumber\\[2ex]
 \mbox{Eq. 46 \cite{alla} }&&  \quad CC_F=5.8 \quad NN_C = - 31.0 
\quad TT_R=14.2 \,,
\nonumber\\[2ex]
\mbox{our result } && \quad CC_F=- 2.83  \quad NN_C = - 42.38 
\quad TT_R= 13.75 \,.
\end{eqnarray} 
~From the results above we infer that the discrepancies mainly occur in
the $C_F^2$ and $C_AC_F$-terms of Eq. (\ref{eq34}).
Substituting $C_F=4/3$ in the expression above we obtain for the coefficient   
of the $(\alpha_s(Q)/\pi)^2$-term the value $-9.49$ instead of $-2.6$
quoted in Eq. (4.7) of \cite{alla} which amounts to a discrepancy of a 
factor of about 3.7. Notice that the bulk of the second order correction
to Eq (\ref{eq34}) is coming from $f_A^{a,(2)}$ in Eq. (\ref{eq33}) which
amounts to 9.216. The remaining part can be traced back to the functions
$f_k^{l,(2)}$ in Eqs. (\ref{eq31}), (\ref{eq32}) leading to the contribution
0.409.\\
Finally we are now also able to present the second order correction for the
shape coefficient (\ref{eq10}) in an analytical way
\begin{eqnarray}
\label{eq38}
a^{\rm H}(Q^2)&=&a^{{\rm H},(0)}(Q^2)\left [ 1 -
\frac{\alpha_s(\mu)}{4\pi} C_F \left \{ 8 \right \} +
\left (\frac{\alpha_s(\mu)}{4\pi}\right )^2 \left \{ C_F^2 \left ( 60 \right )
\right. \right.
\nonumber\\[2ex]
&&\left.\left.+ C_AC_F \left ( \frac{88}{3} \ln\left (\frac{Q^2}{\mu^2}\right ) 
- \frac{1520}{9} \right ) 
+n_fC_F T_F \left(-\frac{32}{3}\ln\left(\frac{Q^2}{\mu^2}\right) 
+ \frac{544}{9} \right ) \right \} \right ] \,,
\end{eqnarray}
which can be written in the same way as has been done for 
$A_{\rm FB}^{\rm H}$ in Eq. (\ref{eq35}) so that we get
\begin{eqnarray}
\label{eq39}
a^{\rm H}(Q^2)
 =  a^{{\rm H},(0)}(Q^2)\left [1 - \frac{\alpha_s(\mu)}{\pi} d_1 
- \left (\frac{\alpha_s(\mu)}{\pi} \right )^2 d_2 \right ] \,.
\end{eqnarray}
We will now discuss the effect of the order $\alpha_s^2$ corrections on
the forward backward symmetry and the shape coefficient for the bottom and 
the charm quark. Further we omit any effect coming from the electroweak sector.
Our results are obtained by choosing the following
parameters (see \cite{caso}). The electroweak constants are: $M_Z = 91.187
~{\rm GeV/c^2}$, $\Gamma_Z = 2.490~{\rm GeV/c^2}$ and $\sin^2 \theta_W =
0.23116$. For the strong parameters we choose : $\Lambda_5^{{\overline {MS}}}
=237~{\rm MeV/c}$ ($n_f=5$) which implies $\alpha_s(M_Z)=0.119$ (two-loop
corrected running coupling constant). Further we take for the renormalization
scale $\mu=Q$ unless mentioned otherwise. Notice that we study 
$A_{\rm FB}^{\rm H}$
and $a^{\rm H}$ for ${\rm H}=c,b$ at the CM energy $Q=M_Z$. For the
heavy flavour masses the following values are adopted : $m_c= 1.50
~{\rm GeV/c^2}$,  $m_b=~4.50~{\rm GeV/c^2}$ and $m_t= 173.8~{\rm GeV/c^2}$. The
results for the bottom quark can be found in table \ref{fig1}. 
The values for $c_2$
are obtained by adding to Eq. (\ref{eq34}) the contributions mentioned
below Eq. (\ref{eq30}). They were denoted by $F_{QCD}^{2-jet}$, 
$F_{QCD}^{3-jet}$ and $F_{QCD}^{\rm F}$ in \cite{alla}.
For the value of the top mass given above we obtain for bottom
production $c_{2,QCD}^{2-jet}=-0.645$,
$c_{2,QCD}^{3-jet}=-0.218$ and $c_{2,QCD}^{\rm F}=0.123$. In the entry
for $c_2$ we have also mentioned between the brackets the result obtained from
the contribution of (\ref{eq34}) which is equal to $c_2=77/8$.
Notice that the second order contributions to the quantities in the tables
are obtained by multiplying $A_{\rm FB}^{(0)}$ at $m=0$ with $c_2$.
Furthermore we have put in the tables the correction in percentages
of the radiatively corrected quantities $A_{\rm FB}^H$, $a^{\rm H}$ with
respect to their zeroth order result.
~From table \ref{fig1} we infer that the order $\alpha_s$ as well as the order 
$\alpha_s^2$ corrections to both $A_{\rm FB}^b$ and $a^b$ are negative. In the
case of the forward-backward asymmetry the QCD corrections are moderate in
particular the second order ones. The latter would be even smaller 
if we had taken the value $c_2=1.9$ quoted in \cite{alla} which is
a factor 4.7 less with respect to our result. In the case of
the shape coefficient the QCD corrections are at least twice as large. The 
latter are reduced if for the reference axis the thrust axis is taken instead 
of the quark axis \cite{dkz}. Notice that the quark axis
has been chosen in our calculation. Further we want to mention that the
zero mass approximation for the second order coefficient $c_2$ is quite 
reasonable. This is revealed by the first order coefficient when the quark mass
is chosen to be zero which leads to the values $c_1=1$ Eq. (\ref{eq34}) and 
$d_1=8/3$ Eq. (\ref{eq38}). In the case of the bottom quark
we observe a deviation of 11 \% for $c_1$ whereas for $d_1$ it amounts to
22 \% which is twice as large. For the charm quark (see table \ref{fig2})
these values become smaller i.e. about 7.5 \%  for both coefficients. If
we expect that the same deviations occur for the coefficients $c_2$ 
Eq. (\ref{eq34})and $d_2$ Eq. (\ref{eq38}) one
gets a reasonable estimate of the theoretical error on the second order
corrections. 
\begin{table}
\begin{center}
\begin{tabular}{|c|c|c|}\hline
\multicolumn{3}{|c|} {$A_{\rm FB}^b$} \\ \hline \hline
        & $m_b=4.50~{\rm GeV/c^2}$  & 
${\overline {m}}_b(M_Z)=2.80~{\rm GeV/c^2}$ 
\\ \hline \hline
$c_1$              & 0.789             & 0.858             \\
$c_2$              & 8.89 (9.63)       & 8.89 (9.63)       \\ 
$A_{\rm FB}^{(0)}$ & 0.1052            & 0.1052            \\
$A_{\rm FB}^{(1)}$ & 0.1020 (-~3.04 \%) & 0.1017 (-~3.33 \%) \\
$A_{\rm FB}^{(2)}$ & 0.1007 (-~4.28 \%) & 0.1004 (-~4.56 \%) \\ \hline \hline
\multicolumn{3}{|c|}{$a^b$} \\ \hline \hline
$d_1$              &  2.08             &  2.33             \\
$d_2$              & 23.0              & 23.0              \\ 
$a^{(0)}$          &  0.994            &  0.998            \\ 
$a^{(1)}$          &  0.916 (-~7.85 \%) &  0.910 (-~8.82 \%) \\
$a^{(2)}$          &  0.883 (-~11.2 \%) &  0.877 (-~12.1 \%) \\ \hline
\end{tabular}
\end{center}
\caption{The forward-backward asymmetry and the shape constant of the bottom
quark.}
\label{fig1}
\end{table}
We also studied the effect of the running quark mass on the forward-backward
asymmetry and the shape coefficient. For this purpose one has to change
the on-mass shell scheme used in Eqs. (\ref{eq16})-(\ref{eq19}), (\ref{eq25})
into the ${\overline {\rm MS}}$-scheme. This can be done by substituting in
all expressions the fixed pole mass m by the running mass ${\overline {m}}
(\mu)$. Moreover one has to add to the first order contributions 
(\ref{eq16})-(\ref{eq19}), (\ref{eq25}) the finite counter term
\begin{eqnarray}
\label{eq40}
\Delta f_k^{l,(1)} = {\overline {m}}(\mu) C_F \left [ 4 - 
3 \ln \left (\frac{{\overline {m}}^2(\mu)}{\mu^2}\right ) \right ]
\left (\frac{d~f_k^{l,(0)}(\rho)}{d~m} \right )_{m={\overline {m}}(\mu)} \,,
\end{eqnarray}
where $\mu$ stands for the mass renormalization scale for which we choose
$\mu=Q$.
Further we adopt the two-loop corrected running mass with the initial
condition ${\overline {m}}(\mu_0)=\mu_0$. Using the relation between
the $\overline{MS}$-mass and the fixed pole mass, as is indicated by the first
factor on the right-hand side in Eq. (\ref{eq40}), we have taken for bottom
production $\mu_0=4.10~{\rm GeV/c^2}$ which corresponds with a pole mass
$m_b=4.50~{\rm GeV/c^2}$. This choice leads to 
$\overline {m}_b(M_Z)=2.80~{\rm GeV/c^2}$ which is 5 \% above the experimental 
value 2.67 ${\rm GeV/c^2}$ measured at LEP \cite{abreu}.
The results are presented in the second column
of table \ref{fig1}. Comparing the latter with the first column we observe
that the QCD corrections become a little bit more negative. The values
of $A_{\rm FB}^{(i)}$ ($i=0,1,2$) decrease a little too. The same features
are also shown by $a^{(i)}$ except for $a^{(0)}$ which slightly increases.

\begin{table}
\begin{center}
\begin{tabular}{|c|c|c|}\hline
\multicolumn{3}{|c|}{$A_{\rm FB}^c$} \\ \hline \hline
        & $m_c=1.50~{\rm GeV/c^2}$  & 
${\overline {m}}_c(M_Z)=0.662~{\rm GeV/c^2}$ 
\\ \hline \hline
$c_1$              &  0.924             &  0.962             \\
$c_2$              & 11.5 (9.63)        & 11.5 (9.63)        \\
$A_{\rm FB}^{(0)}$ &  0.0750            &  0.0750            \\
$A_{\rm FB}^{(1)}$ &  0.0724 (-~3.47 \%) &  0.0723 (-~3.60 \%) \\
$A_{\rm FB}^{(2)}$ &  0.0712 (-~5.07 \%) &  0.0710 (-~5.20 \%) \\ \hline \hline
\multicolumn{3}{|c|}{$a^c$} \\ \hline \hline
$d_1$              &  2.46              &  2.58              \\
$d_2$              & 23.0               & 23.0               \\
$a^{(0)}$          &  1.000             &  1.000             \\
$a^{(1)}$          &  0.907 (-~9.30 \%)  &  0.903 (-~9.70 \%)  \\
$a^{(2)}$          &  0.874 (-~12.6 \%)  &  0.870 (-~13.0 \%)  \\ \hline
\end{tabular}
\end{center}
\caption{The forward-backward asymmetry and the shape constant of the charm
quark.}
\label{fig2}
\end{table}
In table \ref{fig2} we also present results for the charm quark. 
In the case of the running mass we have chosen $\mu_0=1.30~{\rm GeV/c^2}$
so that the pole mass becomes $m_c=1.50~{\rm GeV/c^2}$. This leads to
a value of ${\overline{m}}_c(M_Z)=0.662~{\rm GeV/c^2}$ which is rather
low. Furthermore for charm quark production
the additional contributions become $c_{2,QCD}^{2-jet}=1.359$, 
$c_{2,QCD}^{3-jet}=0.323$ and $c_{2,QCD}^{\rm F}=0.211$. Our result for
$c_2$ is about 2.6 times larger than the value 4.4 obtained in \cite{alla}.
The features are the same as for the bottom quark. The differences between the
numbers in the left-hand and right-hand column in table \ref{fig2} become even
less. The reason
that the running of the mass in the case of the charm and the bottom
quark hardly introduces any effect on the forward-backward asymmetry and
the shape constant can be attributed to the fact that the mass of both quarks
are small with respect to the CM energy so that the phase space is not much 
affected. Moreover, as far as the dynamics is concerned, these quantities are
not proportional to the mass. This is also revealed by the 
constants $c_1$ and $d_1$ which do not deviate very much from their zero
mass values. These arguments do not apply to the top quark except when 
$Q\gg m_t$.
This is shown in table \ref{fig3} where we have studied the above quantities
at a CM energy $Q=500~{\rm GeV/c}$. For the running
mass we have chosen $\mu_0=166.1~{\rm GeV/c^2}$ so that the pole mass
becomes equal to $m_t=173.8~{\rm GeV/c^2}$. This leads to a value
${\overline {m}}_t(Q)=153.5~{\rm GeV/c^2}$. The constants $c_1$ and
$d_1$ in the perturbation series completely differ from the ones given at $m=0$
at which they become 1 and 8/3 respectively.
Therefore the running mass will have a large effects on these constants
which is revealed by the switch of sign in table \ref{fig3}. Hence 
the Born approximations to $A_{\rm FB}^t$ and $a^t$ will change 
while going from the fixed pole mass to the running mass approach.
However the order $\alpha_s$ corrected quantities are less sensitive to
the choice between the running or the fixed pole mass because of the 
compensating
term in Eq. (\ref{eq40}). From the above it is clear that the zero mass
approximation to $c_2$ and $d_2$ makes no sense in case of the top quark
and we have omitted these contributions to $A_{\rm FB}^t$ and $a^t$ in table
\ref{fig3}. Finally we want to comment on the renormalization scale
dependencies of the forward backward asymmetry and the shape constant.
If we vary the scale $\mu$ between $Q/2$ and $2~Q$ the changes in 
$A_{\rm FB}^{(2)}$  are small. It introduces an error of
0.002 for the bottom quark and 0.003 for the charm quark. For $a^{(2)}$ one
can draw the same conclusion and the error becomes 0.005 and 0.007 respectively.
In the case of the top quark a variation in the renormalization scale
makes no sense because of the missing order $\alpha_s^2$ correction. Its
computation for massive quarks will be a enormous enterprise.

\begin{table}
\begin{center}
\begin{tabular}{|c|c|c|}\hline
\multicolumn{3}{|c|}{$A_{\rm FB}^t$} \\ \hline \hline
        & $m_t=173.8~{\rm GeV/c^2}$  & 
${\overline {m}}_t(Q)=153.5~{\rm GeV/c^2}$ \\ \hline \hline
$c_1$              & -~0.757             &  2.11               \\
$A_{\rm FB}^{(0)}$ &  0.407             &  0.455               \\
$A_{\rm FB}^{(1)}$ &  0.417 (2.46 \%)  &  0.426 (-~6.37  \%)   \\ \hline \hline
\multicolumn{3}{|c|}{$a^t$} \\ \hline \hline
$d_1$              & -~0.911             &  5.13               \\
$a^{(0)}$          &  0.406             &  0.514              \\
$a^{(1)}$          &  0.417 ( 2.71 \%)  &  0.435 (-~15.4 \%)   \\ \hline
\end{tabular}
\end{center}
\caption{The forward-backward asymmetry and the shape constant of the top
quark at $Q=500~{\rm GeV/c}$.}
\label{fig3}
\end{table}
Summarizing our findings we have computed the order $\alpha_s^2$
contributions to the forward-backward asymmetry and the shape constant in
an analytical way provided the heavy flavour mass is chosen to be zero.
Further we found a discrepancy with a numerical result calculated earlier
in the literature for $A_{\rm FB}^{{\rm H}(2)}$. The second order corrections
are noticeable. The transition from the fixed pole mass to the running mass
approach does not introduce large changes in the values of $A_{\rm FB}^{\rm H}$
and $a^{\rm H}$ except for the first order constant in the perturbation series
when $H=t$. This indicates that the zero mass approach breaks down
unless $Q\gg m$. Also a variation of the renormalization scale does not lead to
large effects. The latter are almost equal to the differences between the 
results obtained by the fixed pole mass and the running mass approach.\\[5mm]
\noindent
Acknowledgements\\

The authors would like to thank J. Bl\"umlein for reading the manuscript
and giving us some useful remarks. W.L. van Neerven would like to thank 
P.W. Zerwas for discussions concerning
the importance of mass corrections to $A_{\rm FB}^b$ and $a^b$.
This work is supported by the EC-network under contract FMRX-CT98-0194.

\end{document}